\documentclass[fleqn,12pt,twoside]{article}
\usepackage{amssymb,amsmath}
\usepackage[headings]{espcrc1}

\usepackage{graphicx}
\usepackage[figuresright]{rotating}

\newcommand{\order}[1]{\mathcal{O}(#1)}
\newcommand{\calltoall}{\cite{Foley:2005ac}}
\newcommand{\caria}{\cite{Foley:2004jf}}
\newcommand{\cTSI}{\cite{Morningstar:1999dh}}
\newcommand{\cstout}{\cite{Morningstar:2003gk}}
\newcommand{\ctune}{\cite{Morrin:2006tf}}

\title{Charmonium spectral functions in two-flavour QCD}

\author{G.~Aarts\address[Swan]{Department of Physics, University of
        Wales Swansea, Singleton Park, Swansea SA2 8PP, Wales, UK}, 
  C.~R.~Allton\addressmark[Swan], R.~Morrin\address[TCD]{School of
        Mathematics, Trinity College, Dublin 2, Ireland},
        A.~\'O~Cais\addressmark[TCD], M.~B.~Oktay\addressmark[TCD], 
	M.~J.~Peardon\addressmark[TCD],
        J.~I.~Skullerud\addressmark[TCD]\thanks{Speaker}}
       
\runauthor{J.~I.~Skullerud}

\begin{document}

\maketitle
\bibliographystyle{h-elsevier3}

\begin{abstract}
We compute charmonium spectral functions in 2-flavour QCD using the
maximum entropy method and anisotropic lattices.  We find that the
S-waves ($J/\psi$ and $\eta_c$) survive up to temperatures close to
$2T_c$, while the P-waves ($\chi_{c0}$ and $\chi_{c1}$) melt away
below $1.3T_c$.
\end{abstract}

\section{INTRODUCTION}

The fate of charmonium at high temperature has generated much
interest the past 20 years \cite{Matsui:1986dk}.  Contrary to the
original prediction that $J/\psi$ production would be highly
suppressed immediately above the phase transition, recent theoretical
and experimental results indicate a more complicated picture, where
the 1S states may survive up to high temperatures (possibly even with
enhanced yields), while other states disappear earlier.

The properties of hadrons at high temperature are encoded in the
spectral function $\rho_\Gamma$, which can be related to
euclidean-time correlation functions $G_\Gamma$ as follows,
\begin{equation}
G_\Gamma(\tau,\vec{p}) = 
\int_0^\infty\frac{d\omega}{2\pi}\rho_\Gamma(\omega,\vec{p})
\frac{\cosh[\omega(\tau-1/2T)]}{\sinh(\omega/2T)}
\label{eq:spectral}
\end{equation}
where the subscript $\Gamma$ correspond to the different quantum numbers.
Determining the continuous function $\rho(\omega)$ from $G(\tau),
\tau=1\ldots L_{\tau}$, is an ill-posed problem, but it is possible to
infer the most likely $\rho(\omega)$ using the maximum entropy method
(MEM) \cite{Asakawa:2000tr}.  This has been used to study spectral
functions in the quenched approximation
\cite{Asakawa:2003re,Datta:2003ww,Umeda:2002vr}, but substantial
uncertainties still remain.

In order to resolve some of these uncertainties, lattice simulations
with dynamical fermions (2 or 2+1 light flavours) would be highly
desirable.  However, in order to determine spectral functions using
MEM, at least $\order{10}$ independent lattice points are needed in the imaginary time
direction.  At $T\sim2T_c$, this implies a temporal lattice spacing
$a_t\lesssim0.025$fm.  If the spatial lattice spacing $a_s$ were to be
the same, such a simulation would be far too expensive to carry out
with current computing resources.

In order to make the simulation feasible, anisotropic lattices, with
$a_t\ll a_s$, are therefore required.  However, dynamical anisotropic
lattice simulations introduce additional complications not present in
isotropic or quenched anisotropic simulations.  The anisotropic
formulation introduces two additional parameters (the bare quark and
gluon anisotropies), which must be tuned so that the physical
anisotropies are the same for gauge and fermion fields.  In the
presence of dynamical fermions, this requires a simultaneous
two-dimensional tuning, which has been described and carried out in
Ref.~\ctune.

We have previously presented preliminary
results \cite{Morrin:2005zq,Aarts:2005fx}, indicating that S-wave
states survive well into the plasma phase.  These simulations were
carried out before the tuning was completed, and therefore had
systematic uncertainties at the 20\% level.  Here we present results
from a simulation where the anisotropies have been tuned to within
3\%, and study P-waves in addition to S-waves.

\section{RESULTS}

We use the TSI gauge action \cTSI\ and a coarse Wilson, fine
Hamber--Wu fermion action \caria\ with stout-link smearing \cstout\ as
described in \ctune.  The spatial lattice spacing is
$a_s\approx0.18$fm and the anisotropy $\xi=6$, while
$m_\pi/m_\rho=0.54$; our simulation parameters correspond to run 6 in
\ctune.  For the charm quarks we have used bare masses $a_tm=0.08$ and
0.092.  The critical temperature was determined from studying the
variation of the Polyakov loop as a function of $N_\tau$ on
$12^3\times N_\tau$ lattices, giving $N_\tau^c=1/aT_c\approx34$.
500 configurations were generated on
$8^3\times32, 8^3\times24$ and $8^3\times16$ lattices, corresponding
to $T/T_c\approx1.05, 1.4$ and 2.1 respectively.  
Charmonium correlators were computed using all-to-all
propagators \calltoall\ and analysed with Bryan's MEM algorithm
\cite{Bryan:1990} using the free continuum spectral function $\omega^2$ as
default model.

\begin{figure}[t]
\begin{center}
\includegraphics[width=0.45\textwidth]{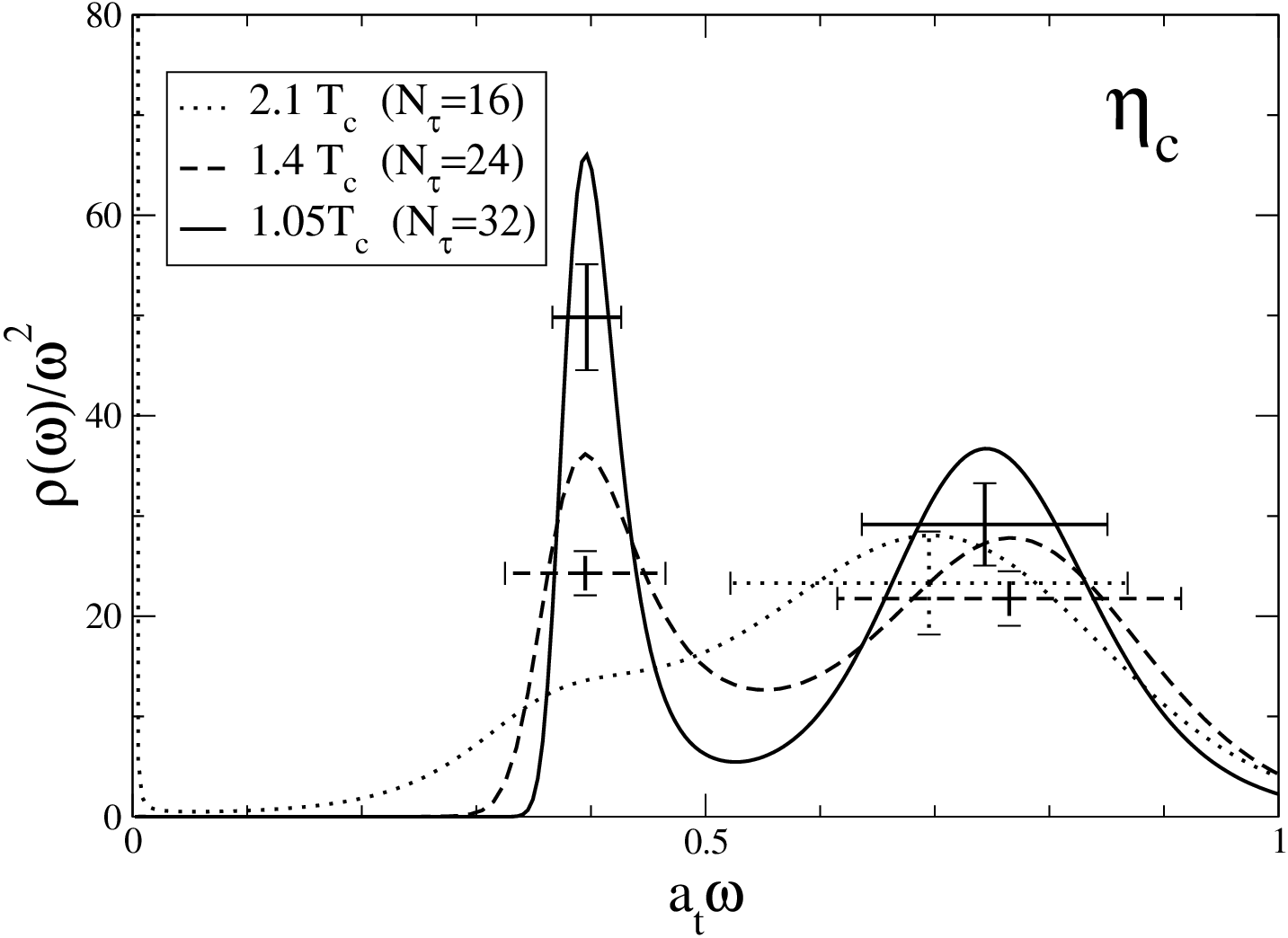}
\includegraphics[width=0.45\textwidth]{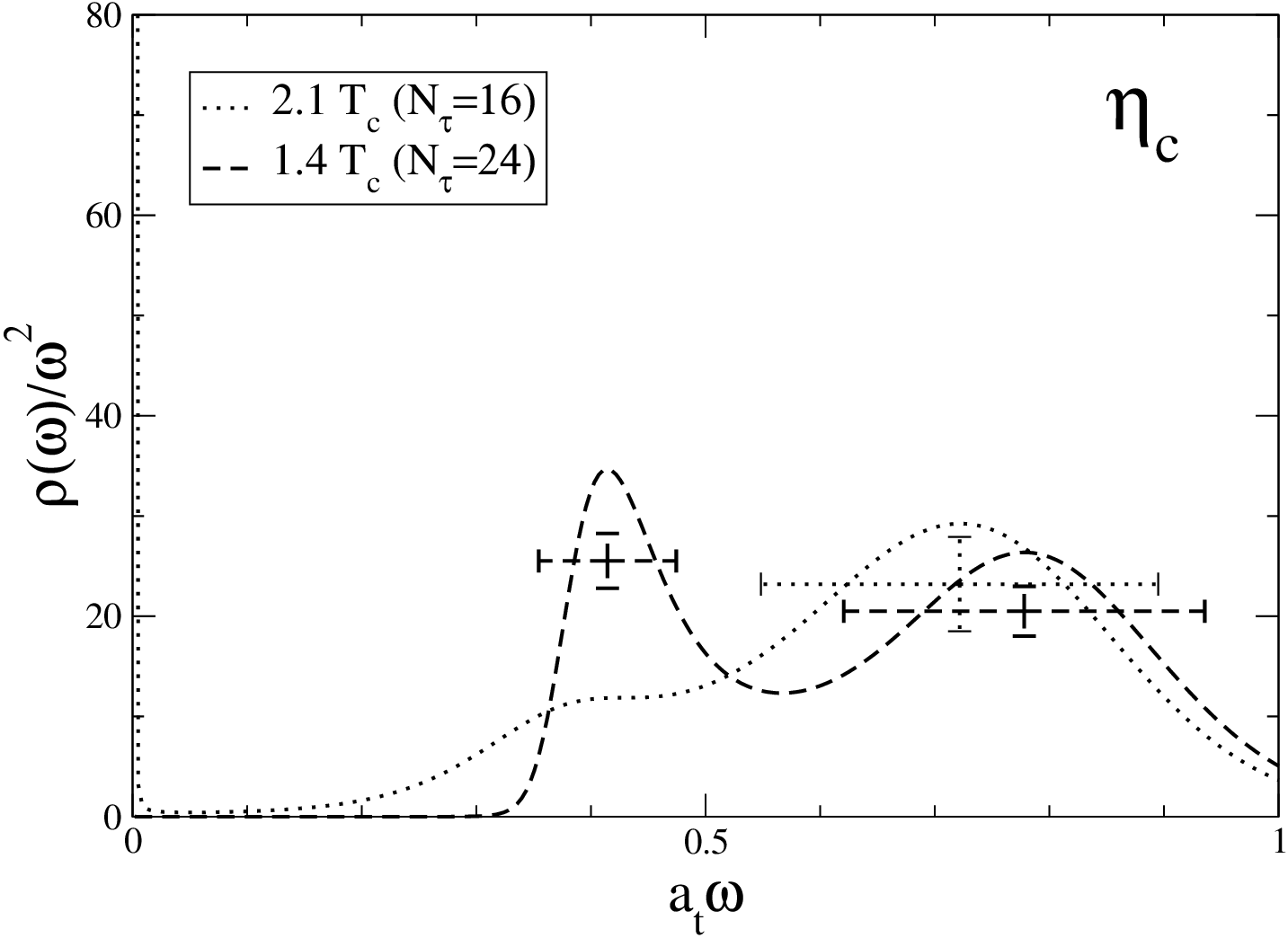}\\
\includegraphics[width=0.45\textwidth]{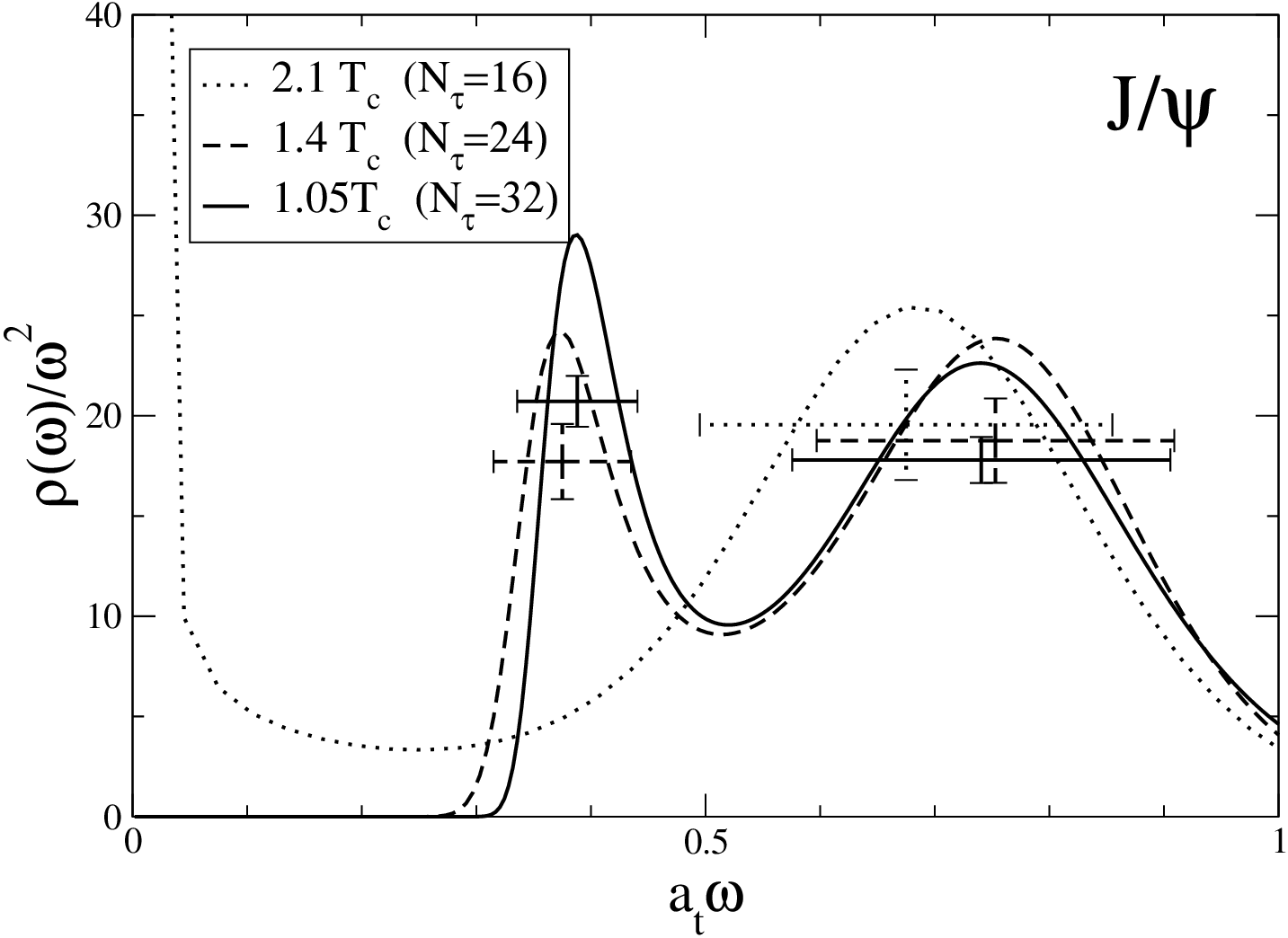}
\includegraphics[width=0.45\textwidth]{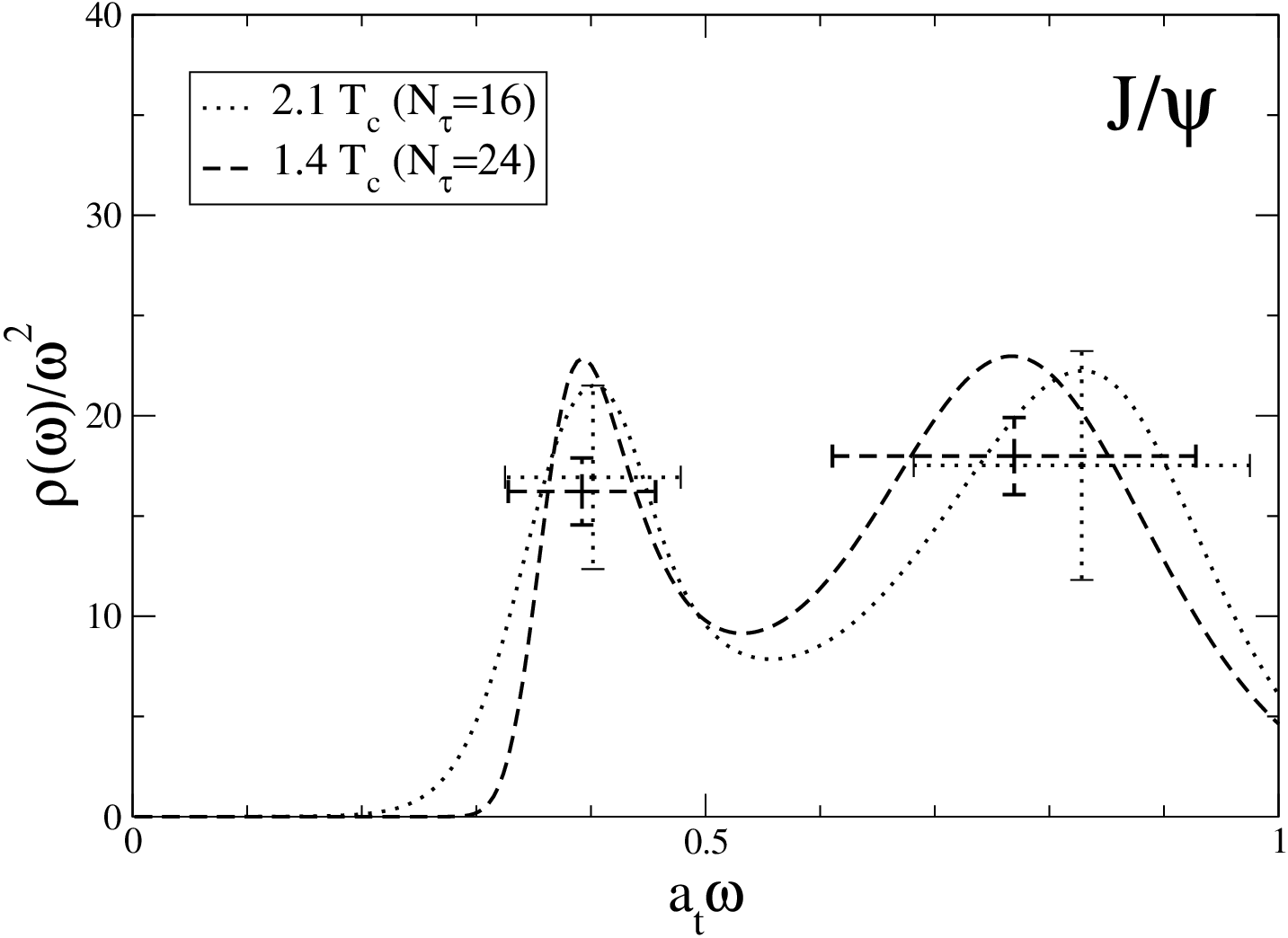}
\caption{Spectral functions for the S-wave charmonium states $\eta_c$
  (top) and $J/\psi$ (bottom), for $a_tm_0=0.080$ (left) and 0.092
  (right).}
\label{fig:swaves}
\end{center}
\end{figure}
Figure~\ref{fig:swaves} show spectral functions for the pseudoscalar
and vector channels, corresponding to the S-wave states $\eta_c$ and
$J/\psi$ respectively.  At the two lower temperatures ($N_\tau=32)$
and $N_\tau=24$) we see two peaks; however, the second (higher) peak
is a lattice artefact which can be observed as a cusp in the free
lattice spectral function \cite{Morrin:2005zq}.  The lower peak
position corresponds to the S-wave ground state masses at zero
temperature, indicating that these states survive more or less
unchanged at least up to $1.4T_c$.  At $T\sim2.1T_c$ the picture is
less clear: for the lighter mass ($m=0.08$) the peak has disappeared
in both channels; while for the heavier mass ($m=0.092$) there is
still a peak in the vector channel, but not in the pseudoscalar.  To
what extent these effects are genuine, i.e.\ whether they show a real
mass dependence and the survival of the vector meson to higher
temperatures than the pseudoscalar, cannot be ascertained at present.

\begin{figure}[t]
\includegraphics[width=0.45\textwidth]{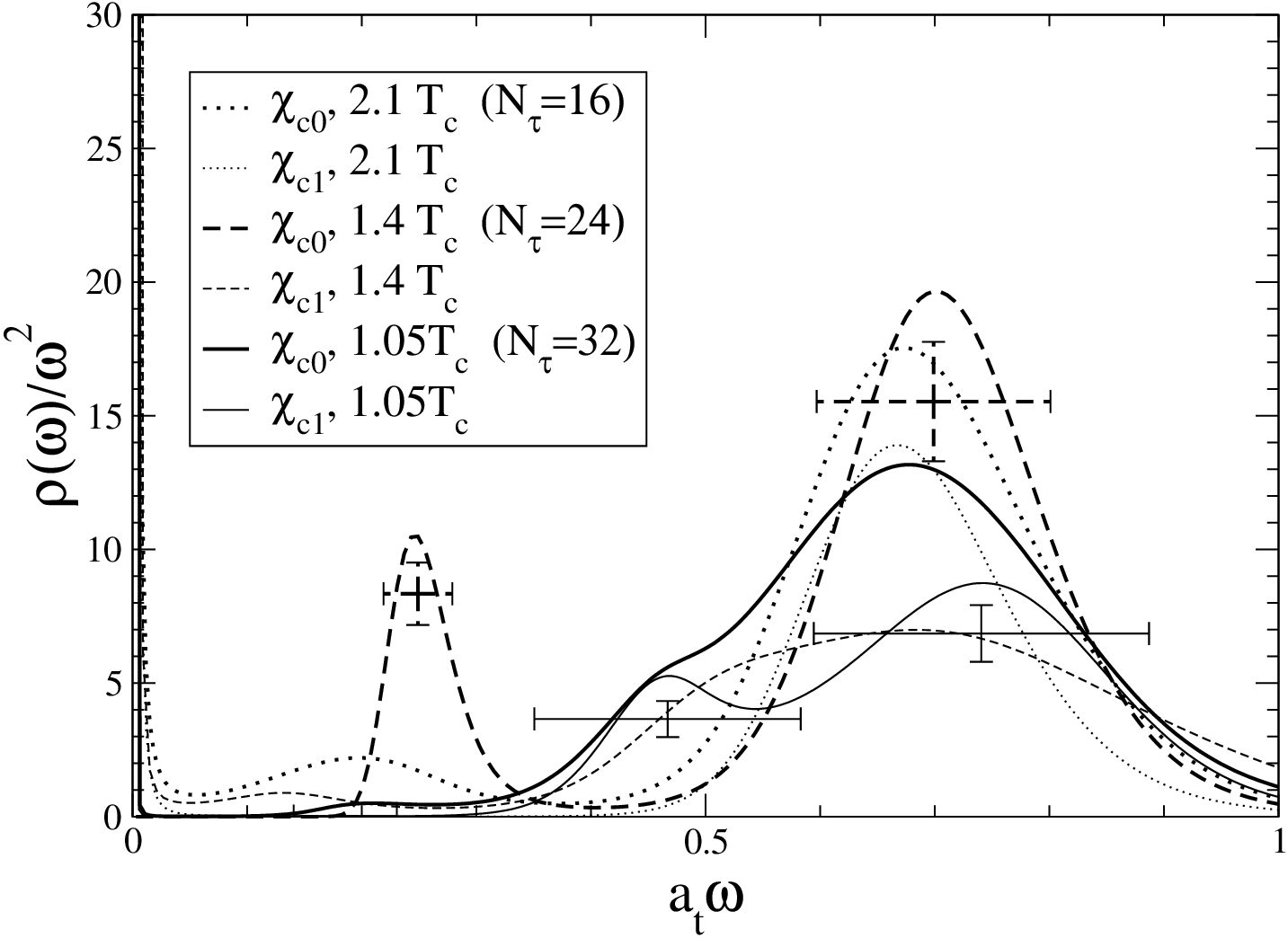}
\includegraphics[width=0.45\textwidth]{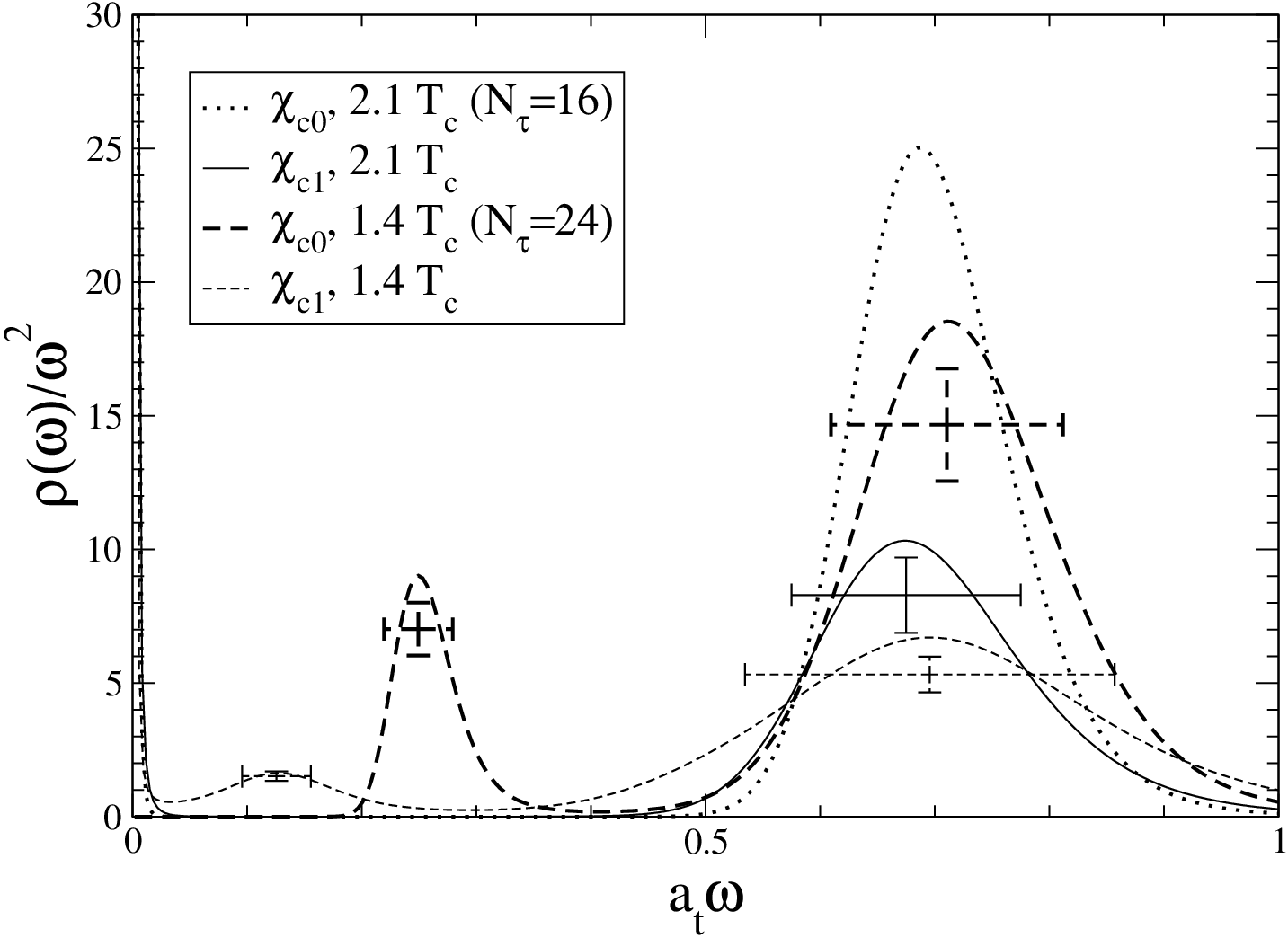}
\caption{Spectral functions for the P-wave charmonium states
  $\chi_{c0}$ and  $\chi_{c1}$, for $a_tm_0=0.080$ (left) and 0.092 (right).
  Thick lines correspond to $\chi_{c0}$, thin lines to $\chi_{c1}$.}
\label{fig:pwaves}
\end{figure}

In fig.~\ref{fig:pwaves} we show spectral functions for the scalar and
axial-vector channels, corresponding to the P-wave states $\chi_{c0}$
and $\chi_{c1}$ respectively.
In the axial-vector channel there is a peak at $a_t\omega\approx0.45$
for the lowest temperature, consistent with the presence of a modified
$\chi_{c1}$ at $1.05T_c$.  There is no discernible corresponding peak
in the scalar channel.  At the higher temperatures the peak has
disappeared, indicating that these states have melted below $1.4T_c$.
The peak in the scalar spectral function at $a_t\omega\approx0.25$ at
$N_\tau=24$ is not clearly understood at present and requires further
investigation.

\section{DISCUSSION}

The main conclusion that can be drawn from this study is that the
S-wave states $J/\psi$ and $\eta_c$ survive virtually unchanged in the
medium above $T_c$, before melting at $T\lesssim2T_c$.  The P-wave
states, on the other hand, disappear shortly above the phase
transition.  Our results thus lend support to the qualitative picture
that has emerged from quenched calculations.
In order to make these results quantitative, a number of outstanding
issues needs to be addressed:
\begin{itemize}
\item The $8^3$ lattice corresponds to a physical volume of
  $(1.4\text{fm})^3$, and finite volume effects are expected to be
  substantial, especially for the P-waves.  We are currently repeating
  the calculation on $12^3$ lattices.
\item A more detailed temperature scan will be necessary to determine
  the melting points of the different states.  This is currently
  underway.  A zero-temperature run is also in progress to provide a
  baseline for comparison.
\item The present analysis has been carried out using only the free
  continuum spectral function as default model.  In order to assess
  systematic uncertainties arising from the MEM, different default
  models will need to be employed.  In particular, the free lattice
  spectral function \cite{Morrin:2005zq} will provide a useful
  comparison.
\end{itemize}
In the final instance, simulations on finer lattices will be necessary
to resolve the remaining uncertainties.  This will however require a
new nonperturbative tuning exercise, and remains a longer-term
prospect.

\section*{Acknowledgments}
This work was supported by the IITAC project, funded by the Irish
Higher Education Authority under PRTLI cycle 3 of the National
Development Plan and funded by IRCSET award SC/03/393Y, SFI grants
04/BRG/P0266 and 04/BRG/P0275.  We are grateful to the Trinity Centre
for High-Performance Computing for their support.

\bibliography{trinlat,hot}
\end{document}